# Enhancement and Inhibition of Stimulated Processes in a Negative Index Cavity

**Domenico de Ceglia[1,2], Marco Centini[3], Antonella D'Orazio[2], Marco De Sario[2], Vincenzo Petruzzelli[2], Francesco Prudenzano[2], Mirko Cappeddu[1,4], Mark J. Bloemer[1], Michael Scalora[1]**

[1]Charles M. Bowden Research Center, AMSRD-AMR-WS-ST,

Research, Development, and Engineering Center, Redstone Arsenal, AL 35898-50003

[2]Dipartimento di Elettrotecnica ed Elettronica, Politecnico di Bari

Via Orabona 4, 70125 Bari, Italy

[3]INFM at Dipartimento di Energetica, Universita di Roma 'La Sapienza'

Via A. Scarpa 16, 00161 Roma, Italy

[4]Universita' degli Studi di Catania, Dipartimento di Ingegneria Elettrica Elettronica e

dei Sistemi, Via A. Doria 6, 95125 Catania – Italy

**ABSTRACT** We study second harmonic generation in a negative index material cavity. The transmission spectrum shows a gap between the electric and magnetic plasma frequencies, so that localized and anti-localized states are allowed at the band edges. The process is made efficient by local phase matching conditions between a forward-propagating pump and a backward-propagating second harmonic signal. By simultaneously exciting the cavity with counter-propagating pulses one is able to enhance or inhibit absorption and stimulated processes. The control parameter is the relative phase difference between the two incident pulses which determines the interference properties of the fields inside the cavity.

After the first theoretical studies on negative-index materials (NIMs)[1], also referred to as left-handed materials (LHMs), the surprising properties of this class of meta-materials[2] have been proposed by Pendry[3] in a thin metal layer acting as a *perfect lens*. Subsequently, this encouraging result has been experimentally observed in metal slabs[4], as well as in composite structures made by loops (or split-ring resonators) and wires.[5,6] Negative refraction, negative phase velocity, and the super-lens effect are remarkable examples of linear phenomena due to a negative refractive index. The inclusion of nonlinear elements within meta-materials, for instance by adding diodes in the split-ring resonators' paths, can induce cubic or quadratic[7,8] nonlinear responses. Therefore, classical nonlinear processes may be reviewed by exploiting the unusual, and sometimes counterintuitive, properties of NIMs. In earlier works, Agranovich et al.[9], and Shadrivov et al.[10] analyzed second harmonic generation (SHG) at an interface between a positive index material (PIM) and a second-order, nonlinear, loss-less NIM. In both papers, it was found that a forward-propagating pump wave tuned in the spectral region where the index of refraction is negative can be phase-matched to a backward-generated second harmonic signal tuned in the positive index region (by forward- or backward-propagation we mean the direction of energy and momentum flow). Efficient SHG has been also reported by Scalora et al.[11] in reflection from an interface between air and a matched NIM. In this case, even though the absorption length was just a few wavelengths, the *local* phase-matching condition $n(\omega) = -n(2\omega)$ provided by the meta-material still ensured significant conversion efficiencies in a process that evolved near the air-NIM interface.

In this Letter we study nonlinear pulse propagation effects and SHG in a quadratic, nonlinear slab of NIM. By engineering the size of split-ring resonators and wires, one is in principle able to design the electric and magnetic plasma frequencies at the desired spectral positions. When both the macroscopic, relative permittivity $\varepsilon(\omega)$ and permeability $\mu(\omega)$ are simultaneously positive or negative, the material allows light propagation, with corresponding positive or negative phase-velocity. On the other hand, light cannot propagate through the material within the frequency range where $\varepsilon(\omega)$ and $\mu(\omega)$ have opposite signs, as the NIM has metallic properties. We consider the Drude model for both $\varepsilon(\omega)$ and $\mu(\omega)$ to take into account material dispersion and causality, so that

$$\varepsilon(\omega) = \varepsilon_\infty - \frac{\omega_e^2}{\omega^2 + j\omega\gamma_e}$$
$$\mu(\omega) = \mu_\infty - \frac{\omega_m^2}{\omega^2 + j\omega\gamma_m}$$
(1)

where $\varepsilon_\infty$ and $\mu_\infty$ are the permittivity and permeability at high frequencies, $\omega_e$ and $\omega_m$ are the electric and magnetic plasma frequencies, $\gamma_e$ and $\gamma_m$ are the damping factors, which are related to the extinction coefficient of the medium. For simplicity, the cavity we analyze is surrounded by air; moreover, the electric and magnetic plasma frequencies do not coincide, so that $\omega_e < \omega_m$, and the medium is not matched to the surrounding air. With these parameters, an *intrinsic* gap is created between these two frequencies, so that the material displays NIM behavior for $\omega < \omega_e$, and a positive index response for $\omega > \omega_m$. As pointed out by D'Aguanno et al.[12,13], such a structure localizes the electric field of a wave tuned at the lower band edge, and anti-localizes the electric field of a signal tuned at the upper band edge. As a result, significant lowering of the group velocity and large enhancement of the density of modes can be achieved at the band edge using a NIM slab just a few wavelengths thick. This behavior reveals that such a NIM cavity reproduces mechanisms analogous to band edge effects displayed by a one-dimensional photonic crystal, and so it is suitable to enhance nonlinear processes.

In the paper by Centini et al.[14], the dynamics of counter-propagating pulses through a one-dimensional photonic crystal was analyzed, and it was shown that SHG efficiency may be enhanced or suppressed by changing the relative phase difference between the coherent, incident pulses. This effect is due to the phase-dependent interference mechanism that occurs between the two pulses tuned at the band edge, leading to dynamic and coherent control of the *effective* density of modes that regulates stimulated processes. In this paper we thus analyze pulsed SHG in a NIM cavity and exploit the local phase-matching conditions that exist inside the cavity. We also demonstrate the modulation of SHG efficiency by using coherent, counter-propagating pulses. Furthermore, we observe that the inhibition of SHG efficiency is accompanied by a drastic reduction of absorption losses in the cavity.

We solve Maxwell's equations in the time domain using two different techniques: (i) the finite difference time domain (FDTD) scheme described by Ziolkowsky and Heyman[15], supplemented by the introduction of nonlinear magnetic and electric polarization terms; and (ii) the method described in ref. 11 by Scalora et al., which utilizes a fast Fourier transform pulse propagation algorithm. While both give nearly identical results, here we describe the FDTD approach. The following equations are relative to the propagation of pulses along the z direction:

$$\frac{\partial E_x}{\partial t} = -\frac{1}{\varepsilon_0}\left[\frac{\partial H_y}{\partial z} + J_x\right] - \chi_e^{(2)}\frac{\partial E_x^2}{\partial t^2}$$

$$\frac{\partial J_x}{\partial t} + \gamma_e J_x = \varepsilon_0 \omega_e^2 E_x$$

$$\frac{\partial H_y}{\partial t} = -\frac{1}{\mu_0}\left[\frac{\partial E_x}{\partial z} + K_y\right] - \chi_m^{(2)}\frac{\partial H_y^2}{\partial t^2} \quad (2)$$

$$\frac{\partial K_y}{\partial t} + \gamma_m K_y = \mu_0 \omega_m^2 H_y$$

where $J_x = \partial P_x/\partial t$ e $K_y = \partial M_y/\partial t$ are the electric and magnetic currents, $P_x$ and $M_y$ are respectively the polarization and the magnetization, $E_x$ and $H_y$ are the electric and magnetic fields. This set of equations is numerically integrated by using finite differences. The temporal step is $dt=0.9dz$ in order to insure the stability of the method. Second-order, analytical absorbing boundary conditions are adopted at the edges of the simulation grid.[16] The width of the cavity is L=3.85 μm; we normalize the angular frequencies by assuming a reference frequency $\omega_0$ corresponding to the wavelength $\lambda_0 = 1\mu m$. We also assume that $\tilde{\omega}_m = 1$, $\tilde{\omega}_e = 0.528$, and $\tilde{\gamma}_e = \tilde{\gamma}_m = 10^{-4}$. The linear transmission and the refractive index (real and imaginary parts) of the NIM are depicted in Fig.1. The pump signal is tuned at the lower band edge at $\tilde{\omega} = \tilde{\omega}_F = 0.522$, close to the electric plasma frequency, so that an exceptionally high localization of the electric field and an anti-localization of the magnetic field occur. The Q-factor of the cavity, measured at the lower band edge resonance, is about 400. In contrast, waves tuned to the transmission resonances in the neighborhood of the magnetic plasma frequency display the opposite picture, i.e. a strong magnetic field localization and an anti-localization of the electric field. The structure is designed in a way that the second harmonic signal is tuned to the second resonance above the high-frequency band edge, so that the anti-localization of the electric field does not negatively affect the second harmonic (SH) conversion efficiency. Moreover, the particular phase-matching condition $n(\tilde{\omega}_F) = -n(\tilde{\omega}_{SH})$ achieved at $\tilde{\omega}_F = 0.522$ implies that the forward-propagating signal at the fundamental frequency, which experiences negative refraction, is phase-matched to the backward-propagating signal at the second harmonic frequency, as described in ref. 11. Even with a relatively small damping coefficient, the high cavity-Q causes linear absorption losses to reach nearly 20% at the band edge.

In Fig. 2 we show the SHG efficiency as a function of pump pulse duration. Using nonlinear coefficients of $\chi_e^{(2)} = 25\,pm/V$ and $\chi_m^{(2)} = 0$, and an input peak power $P_{in} = 500MW/cm^2$ for the pump pulse, we observe a maximum conversion efficiency of approximately 7% for pulses long enough to be completely coupled with the narrow, band edge resonance at $\tilde{\omega} = 0.522$. The conversion efficiency is calculated by considering the total amount of SH radiated energy, that in this case is nearly equally divided between the forward and backward directions.

The use of coherent, counter-propagating pulses, as described in ref. 14, can increase or decrease the *effective* transit time through the cavity, so that one is able to coherently control the *effective* group velocity and the interaction time between the electromagnetic field and the structure. In Fig. 3 we plot the delay (or transit) time, i.e. the time needed by the peak of the pulse to traverse the entire cavity, as a function of the relative phase difference $\Delta\varphi$ between two coherent, counter-propagating pulses impinging onto the NIM with the same peak power, both tuned to the band edge resonance. The interference between the pulses in the structure is constructive when $\Delta\varphi = 0$, and the delay of both left and right radiated pulses is maximized to approximately double the delay accumulated by a single pulse. Under these operational conditions, the localization of the electric field is maximized, and the *effective* group velocity is reduced. Now, if $\Delta\varphi = 180°$, the two pulses emerging from the cavity undergo a delay that is almost one order of magnitude smaller than the maximum delay value, and the total electromagnetic field becomes anti-localized.

Finally, we analyze the nonlinear interaction under counter-propagating pulses regime; in this excitation scheme, each pulse carries a peak power equal to $P_{in}/2$. The SHG conversion efficiency and the absorption are plotted in Fig. 4. We stress the two phenomena related to the modulation of the relative phase difference: (i) for $\Delta\varphi = 0$, SHG is enhanced by a factor of ~4 with respect to the efficiency achieved by a single pump pulse; for $\Delta\varphi = \pi$ we see a drastic inhibition of SHG as a consequence of the reduced density of modes and anti-localization of the light field; (ii) if we define the absorption losses in the cavity as $A=1-T-R$, where $T$ is the total (fundamental and second harmonic) energy flowing in the forward direction, normalized with respect to $P_{in}$, and $R$ the total, normalized energy flowing in the backward direction, the value of $A$ is then drastically reduced when $\Delta\varphi = 180°$. This effect is due primarily to the reduced interaction time in the cavity (or increase of *effective* the group velocity), so that the pulses interfere destructively giving rise to a reduced density of modes when they reach the center of the structure. Under the circumstances, we find that while each individual pulse transits through the cavity in its pre-established direction, they do not interact with the meta-material.

In conclusion, we have studied the SHG process inside a NIM cavity, under condition of local phase matching between the pump and the second harmonic signal. Given an input power of $\sim P_{in} = 500 MW/cm^2$, $\chi_e^{(2)} = 25 pm/V$, and $\chi_m^{(2)} = 0$, we estimate a conversion efficiency of ~7%. We further remark that it is possible to inhibit or enhance stimulated processes, and to control absorption losses thanks to the interference of coherent, counter-propagating pulses. The relative phase of the incident pulses acts as the control mechanism of these phenomena.

D. de Ceglia thanks US Army for partial financial supports.

**Figure captions**

**Fig. 1.** Transmission spectrum of a 3.85-micron thick NIM as a function of the normalized angular frequency $\tilde{\omega}$. The normalized, electronic and magnetic plasma frequencies are respectively $\tilde{\omega}_e = 0.528$ and $\tilde{\omega}_m = 1$; the normalized damping factors are $\tilde{\gamma}_e = \tilde{\gamma}_m = 10^{-4}$. The dashed and dotted curves are respectively the real and imaginary parts of the refractive index in the cavity. At the mid-gap frequency the transmittivity is of the order of $10^{-10}$ and the real part of the refractive index is about $10^{-6}$. In the tuning configuration indicated by the arrows, the real part of the refractive index satisfies the phase-matching condition $n(\tilde{\omega}_F) = -n(\tilde{\omega}_{SH})$, as shown by the two circles.

**Fig. 2.** Overall SHG conversion efficiency vs. pulse duration of a pump input with a peak power $P_{in} = 500 MW/cm^2$. The efficiency shows saturation around 7% for pulse durations longer than 16ps. In the calculation we assume $\chi_e^{(2)} = 25 pm/V$ and $\chi_m^{(2)} = 0$.

**Fig. 3.** Delay accumulated in the cavity for reflected (backward) and transmitted (forward) pulses as a function of the relative phase difference between the two coherent, incident pulses. Pulse duration is ~3.5ps in all cases. The inset shows the coherent, counter-propagating excitation scheme of the cavity.

**Fig. 4.** SHG efficiency (curve with triangles) and overall energy at the fundamental frequency (curve marked by squares) as functions of the relative, input phase difference. The input pump pulses have a peak power of $P_{in}/2 = 250 MW/cm^2$, and their duration is ~3.5ps. The conversion efficiency of single pulse incident onto the structure is ~5% (dashed line). The absorption reaches a maximum of ~20% for $\Delta\varphi = 0$ and it is virtually suppressed, along with the SHG process, for $\Delta\varphi = 180°$.

Fig.1

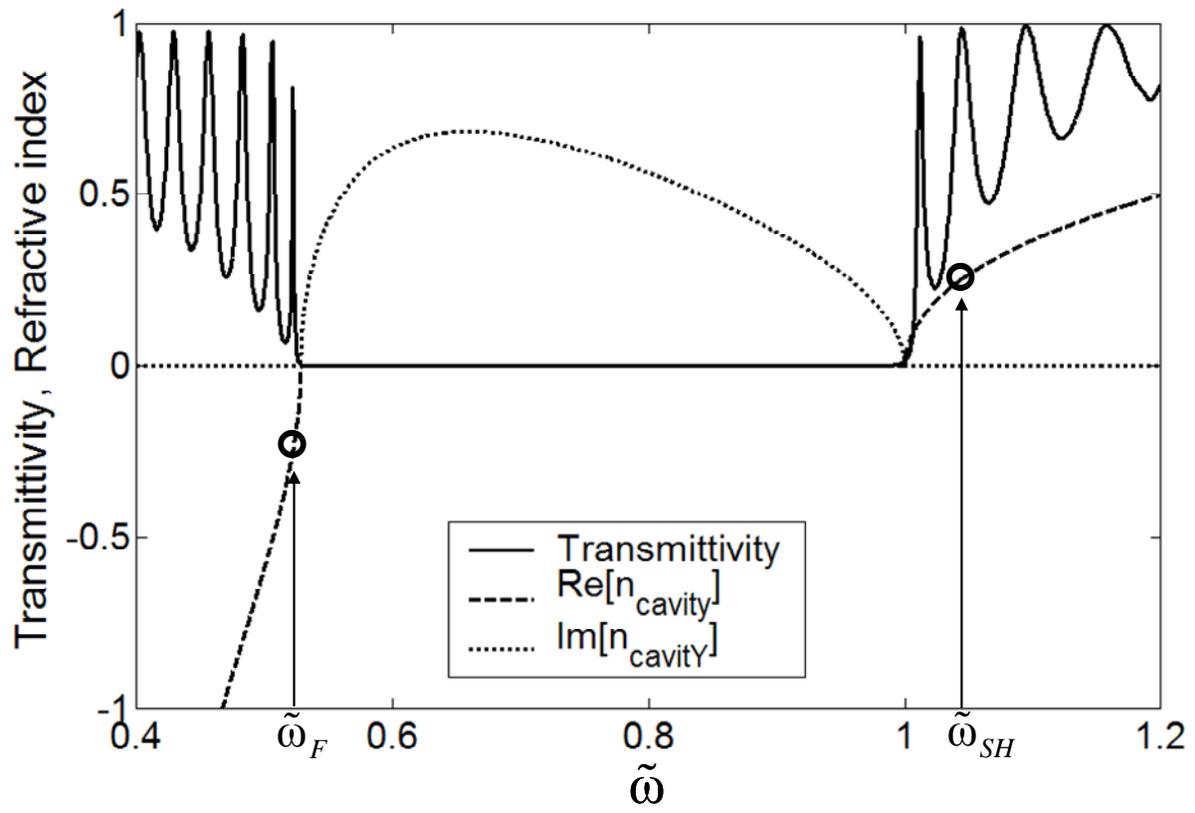

Fig.2

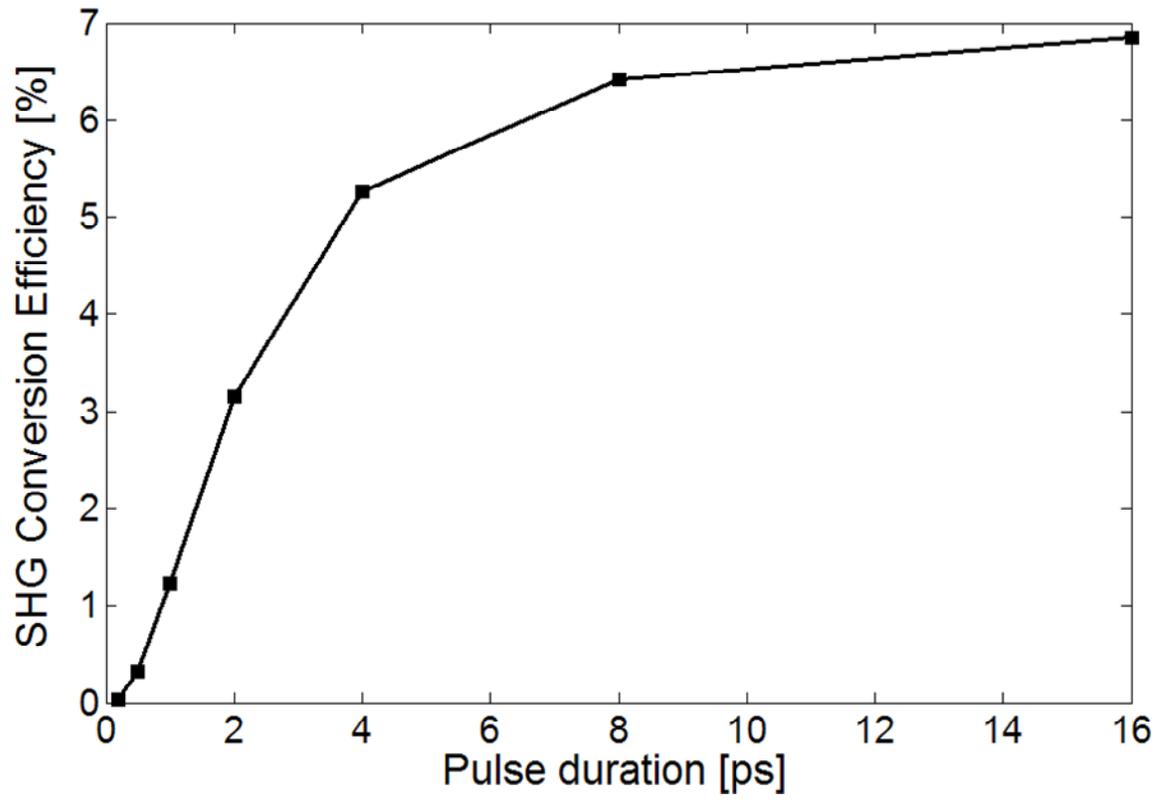

Fig.3

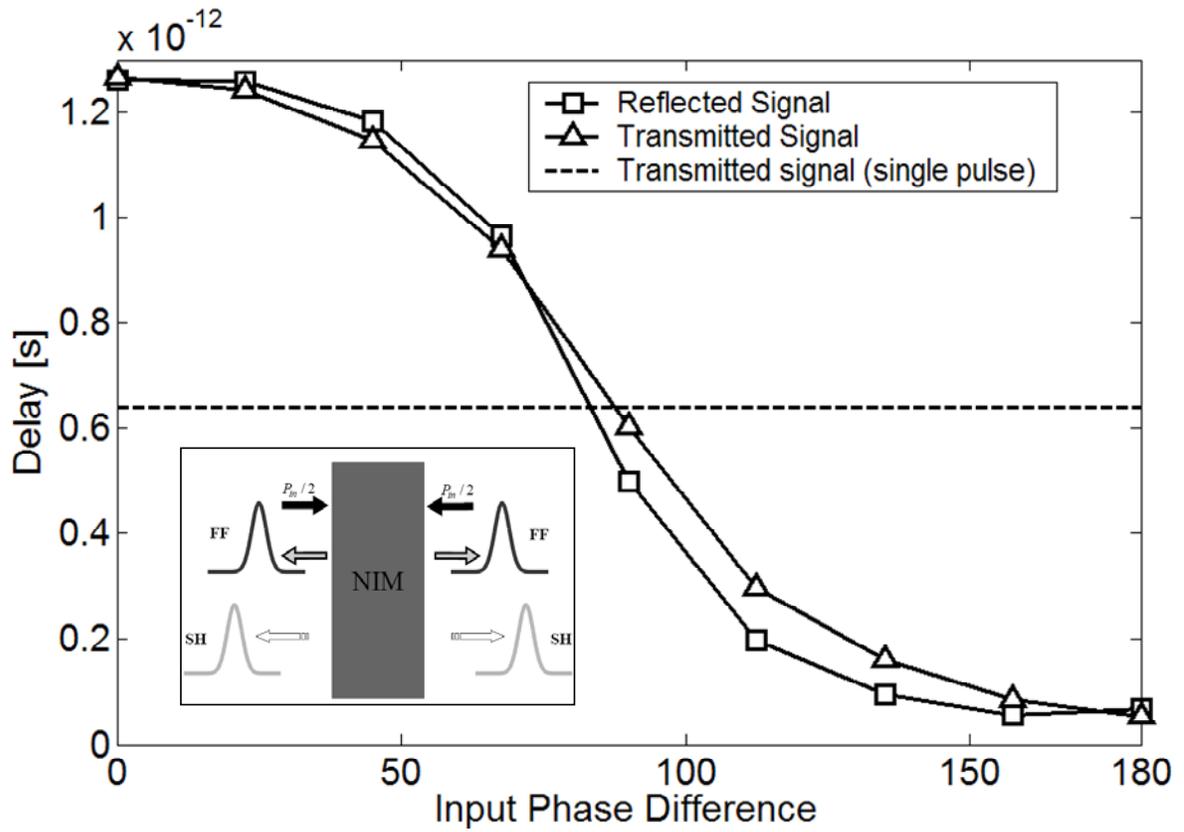

Fig.4

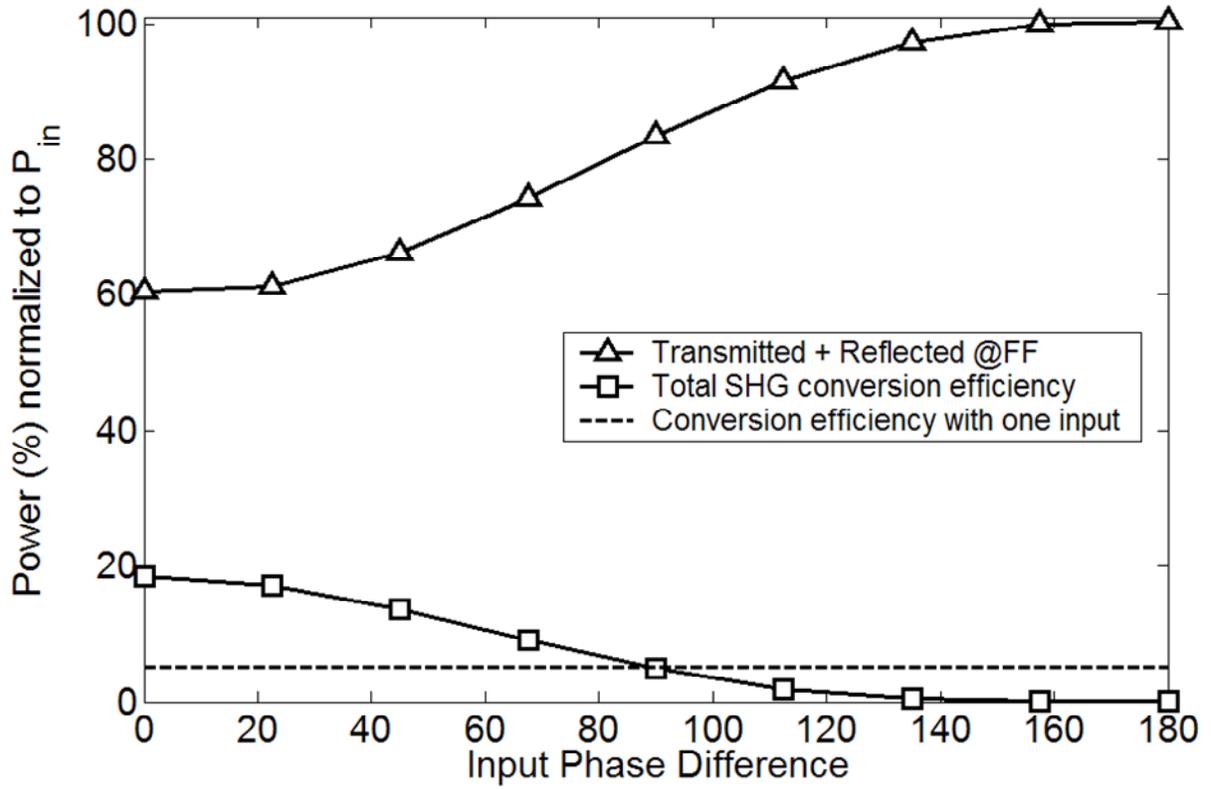